\documentclass[aps,prd,reprint,balancelastpage,nofootinbib,preprintnumbers,amsmath,amssymb,superscriptaddress,longbibliography]{revtex4-1}
\usepackage[
colorlinks=true,        
citecolor=blue,         
linkcolor=blue,         
urlcolor=blue           
]{hyperref}             
\usepackage{bm}         
\usepackage{ragged2e}
\usepackage{xcolor}     
\usepackage{orcidlink}
\usepackage{units}
\usepackage{graphicx}
\usepackage{subcaption}

\usepackage{caption}
\newcommand{\nc}{\newcommand*}

\nc{\Eq}[1]{Eq.~\eqref{#1}}     
\nc{\Fig}[1]{Fig.~\ref{#1}}     
\nc{\Table}[1]{Table~\ref{#1}}  
\nc{\Sec}[1]{Sec.~\ref{#1}}     
\def\({\left(}
\def\){\right)}
\def\[{\left[}
\def\]{\right]}
\def\e{\begin{equation}}
\def\q{\end{equation}}
\def\m{\begin{eqnarray}}
\def\n{\end{eqnarray}}

\usepackage{xcolor}          
\usepackage[normalem]{ulem}  

\usepackage[draft]{changes}

\setaddedmarkup{\textcolor{blue}{#1}}

\setdeletedmarkup{\textcolor{red}{\sout{#1}}}

\begin{document}

\title{Constraining the gravitational-wave emission of core-collapse supernovae with ground-based detectors}


\author{Jingwang Diao\orcidlink{0000-0002-3467-3102}}
\affiliation{Institute for Frontier in Astronomy and Astrophysics and Faculty of Arts and Sciences, Beijing Normal University, Zhuhai 519087, China}
\affiliation{School of Physics and Astronomy, Beijing Normal University, Beijing 100875, China}

\author{Xingjiang Zhu\orcidlink{0000-0001-7049-6468}}
\email{zhuxj@bnu.edu.cn}
\affiliation{Institute for Frontier in Astronomy and Astrophysics \& Faculty of Arts and Sciences, Beijing Normal University, Zhuhai 519087, China}

\begin{abstract}
A gravitational-wave background (GWB) arising from the superposition of numerous unresolved gravitational-wave signals has yet to be detected. 
Potential contributing sources to such a background include compact binary coalescences (CBCs) and core-collapse supernovae (CCSNe).
In this work, we place upper limits on the gravitational-wave energy emitted by CCSNe using cross-correlation measurements made with Advanced LIGO and Advanced Virgo detectors during their third observing run.
Specifically, we obtain a $95\%$ credibility upper limit of $0.01~ {M_\odot c^2}$ while accounting for the contribution from CBC sources to a GWB. This result improves on previous constraint obtained from initial LIGO data by approximately 2 orders of magnitude.
We also explore the detection prospects of third-generation ground-based detectors such as the Einstein Telescope and Cosmic Explorer for both individual CCSNe events and the GWB.
Our results show that single events are likely to be detected prior to the GWB.
\end{abstract}
\maketitle

\section{Introduction} 
Since the first discovery of gravitational waves (GWs) a decade ago~\cite{LIGOScientific:2016aoc}, numerous GW events have been detected, with 218 confirmed events reported in version 4.0 of the Gravitational-Wave Transient Catalog (GWTC-4.0)~\cite{LIGOScientific:2025slb}. However, the signals detected by ground-based interferometers such as LIGO \cite{LIGOScientific:2014pky}, Virgo \cite{VIRGO:2014yos}, and KAGRA \cite{KAGRA:2018plz} represent only a small fraction of the GW sources in the Universe. A substantial number of GW signals remain undetected, including unresolved distant compact binary coalescences (CBCs)~\cite{Zhu:2011bd,Marassi:2011si,Wu:2011ac}, core-collapse supernovae (CCSNe)~\cite{Buonanno:2004tp,Sandick:2006sm}, rotating neutron stars~\cite{Ferrari:1998jf,Regimbau:2001kx,Zhu:2011pt,Lasky:2013jfa}, cosmic strings~\cite{Kibble:1976sj,Sarangi:2002yt,Damour:2004kw,LIGOScientific:2017ikf}, early-Universe phase transitions~\cite{Lopez:2013mqa,Dev:2016feu,Marzola:2017jzl}, etc. The collective contribution from these unresolved sources constitutes a gravitational-wave background (GWB).

During the final stages of stellar evolution, stars with masses exceeding $\sim 8~{M_\odot}$ undergo catastrophic gravitational collapse, resulting in the formation of either a neutron star or a black hole~\cite{Burrows:1995ww,Kotake:2005zn,Janka:2012wk}.
Throughout the collapse process, GWs are expected to be emitted. In 2021, \citet{Szczepanczyk:2021bka} compiled 82 waveforms from multidimensional CCSNe simulations, with GW energies ranging from $10^{-11}$ to $10^{-6}~{M_\odot c^2}$. More recently, \citet{Vartanyan:2023sxm} and \citet{Choi:2024irp} performed three-dimensional radiation-hydrodynamic simulations of nonrotating progenitors and found emitted GW energies spanning approximately $10^{-11}$ - $10^{-8}~{M_\odot c^2}$. In addition to these numerical simulations, several phenomenological models, such as the long-lasting bar-mode instability~\cite{Ott2010} and torus fragmentation instability~\cite{Piro2007}, predict stronger GW emission in the range of $10^{-4}$ - $10^{-2}~{M_\odot c^2}$~\cite{LIGOScientific:2016jvu}. Despite extensive searches, no CCSN-associated GW signals have been detected to date. The main challenge might be the intrinsically weak GW emission from CCSNe, making detection feasible only for nearby events~\cite{Gossan:2015xda,Szczepanczyk:2021bka,KAGRA:2021tnv}. Current-generation detectors are primarily sensitive to CCSNe within the Milky Way, yet the Galactic CCSNe rate is estimated to be only one or two events per century~\cite{Bergh:1991ek,Cappellaro:1993ns,Tammann:1994ev,Diehl:2006cf,Li:2010kc,Adams:2013ana}. 

In 2016, LIGO conducted a search for GWs from two CCSNe events—SN~2007gr and SN~2011dh. Although no signals were detected, limits were placed on the GW emission using sine-Gaussian waveform models. Based on the known distances to these supernovae, the estimated upper limits range from $0.1~{M_\odot c^2}$ at low frequencies to $10~{M_\odot c^2}$ at frequencies above 1~kHz~\cite{LIGOScientific:2016jvu}. During the first and second observing runs of Advanced LIGO and Advanced Virgo, additional CCSNe-targeted GW searches were conducted. Though no significant signals were found, upper limits were placed on the GW energy using \textit{ad hoc} harmonic waveforms windowed with Gaussian envelopes: $E_\text{GW} < 4.27 \times 10^{-4}~{M_\odot c^2}$ at 235~Hz and $E_\text{GW} < 1.28 \times 10^{-1}~{M_\odot c^2}$ at 1304~Hz~\cite{Abbott2020}. More recently, the LIGO–Virgo–KAGRA Collaboration reported on GW searches targeting CCSNe, including SN~2023ixf. No GW signals were detected. Using an ellipsoidal rotating protoneutron star model, the search sensitivity reached $E_\text{GW} \sim 10^{-4}~{M_\odot} c^2$ for emission at 82~Hz~\cite{LIGOScientific:2024jxh}.

Bcause of the incomplete understanding of the CCSNe explosion mechanisms, current physical models remain highly uncertain. Nevertheless, GWs from CCSNe are expected to lie in the frequency range from tens of hertz to several kilohertz, and the entire cosmological CCSNe population could contribute significantly to the GWB. \citet{Zhu:2010af} utilized upper limits on the GWB energy density from initial LIGO data to constrain the average GW emission per CCSNe, placing bounds of $\sim 1\,M_\odot c^2$. Building on this work, we use data from the third observing run (O3) of LIGO and Virgo~\cite{LVK-O3-SGWB} to place upper limits on the GWB produced by CCSNe and thereby constrain the average GW energy. Furthermore, given the non-negligible contribution of CBCs to the GWB, we extend our analysis to include CBCs alongside CCSNe. Finally, we forecast the sensitivity of third-generation GW detectors to both individual CCSNe and their collective GWB signal.

The structure of this paper is as follows. In Sec.~\ref{sec:II}, we describe the GWB model of CCSNe and CBCs. In Sec.~\ref{sec:III}, we use the O3 cross-correlation observations to place upper limits on $E_{\text{GW}}$. We discuss the detectability of individual CCSNe events and the GWB using future GW detectors in Sec.~\ref{sec:IV}. Finally, Sec.~\ref{sec:V} summarizes our findings and conclusions.

\section{ \label{sec:II} The Gravitational-wave background model}

The energy density of the GWB generated by astrophysical sources can be expressed as~\cite{Zhu:2010af}
\begin{equation}
\Omega_{\text{GW}}(f) = \frac{1}{\rho_c c} f \int F_{\rm flux}(f, z) \frac{\mathrm{d}R(z)}{\mathrm{d}z} \, \mathrm{d}z,
\label{eq:omega_gw}
\end{equation}
where $\rho_c = \frac{3 H_0^2 c^2}{8\pi G}$ is the critical energy density, and the integrated flux at the observed frequency $f$ is defined as
\begin{equation}
F_{\rm flux}(f,z) = \frac{1}{4 \pi D_{L}(z)^2} \frac{\mathrm{d}E_{\text{GW}}}{\mathrm{d}f_s}  (1+z)^2 .
\end{equation}
Here, $D_{L}(z) = \frac{c(1+z)}{H_0} \int \frac{\mathrm{d}z}{E(\Omega, z)}$ is the luminosity distance, with $E(\Omega,z)=\sqrt{\Omega_{\Lambda}+\Omega_{m}(1+z)^3}$. Throughout this work, we adopt the cosmological parameters  $H_0 = 67.4\,\mathrm{km\,s^{-1}\,Mpc^{-1}}$, and $\Omega_{m} = 1 - \Omega_{\Lambda} = 0.315$~\cite{Planck:2018vyg}. The term $\mathrm{d}E_{\text{GW}}/\mathrm{d}f_s$ represents the GW energy spectrum emitted by a single source, with $f_s = f(1+z)$ being the frequency in the source frame.

Following Ref.~\cite{Zhu:2010af}, we adopt a Gaussian model to describe the average energy spectrum of CCSNe,
\begin{equation}
\frac{\mathrm{d}E_{\text{GW}}}{\mathrm{d}f_s} = A \exp\left[ - \frac{(f_s - f_{\text{peak}})^2}{2 \Delta^2} \right],
\label{eq:energy_spectrum}
\end{equation}
where $A$ is the normalization amplitude, $f_{\text{peak}}$ is the spectral peak frequency, and $\Delta$ denotes the bandwidth. In this work, $f_{\text{peak}}$ and $\Delta$ are source-frame parameters and are assumed to be independent of redshift.

The differential source event rate $\mathrm{d}R(z)/\mathrm{d}z$ in Eq.~\eqref{eq:omega_gw} is given by
\begin{equation}
\frac{\mathrm{d}R(z)}{\mathrm{d}z} = \frac{\dot\rho_*(z)}{1+z} \frac{\mathrm{d}V}{\mathrm{d}z}(z) \int \Phi(m) \, \mathrm{d}m,
\label{eq-rate}
\end{equation}
where $\dot\rho_*(z)$ is the cosmic star formation rate per comoving volume in units of $M_\odot\,\mathrm{Mpc}^{-3}\,\mathrm{yr}^{-1}$. We adopt the parametrized form of the $\dot\rho_*(z)$ from Ref.~\cite{Madau:2014bja}. Here, $\mathrm{d}V/\mathrm{d}z$ is the comoving volume element, and $\Phi(m)$ is the stellar initial mass function (IMF), for which we use the modified Salpeter A form~\cite{Baldry:2003xi}. We assume that each CCSN event arises from a progenitor with initial mass between 8~$M_\odot$ and 100~$M_\odot$~\cite{Smartt:2009zr}. We note that the progenitor mass range or the choice of IMF enters the GWB model through the overall event-rate normalization in Eq.~\eqref{eq-rate}. This normalization scales the amplitude of $\Omega_{\rm CCSNe}(f)$ and is therefore degenerate with the spectrum normalization parameter $A$. Since our analysis marginalizes over $A$ over a broad prior range, our results are insensitive to the IMF model or CCSNe progenitor mass range.
The true mass range for successful supernova explosions might be narrower than the one adopted here. For example, stars above $\sim 40\,M_\odot$ may collapse directly to black holes, without producing a supernova. However, for the IMF adopted here, changing the mass integral from $8$--$100\,M_\odot$ to $8$--$40\,M_\odot$ reduces the rate normalization by only $8\%$. Therefore, we ignore the uncertainty associated with IMF and CCSNe progenitor mass range in this work.
We note, however, that although the progenitor mass range is degenerate with the normalization
$A$, this parameter directly affects the inferred total GW energy, and the resulting constraints implicitly depend on the assumed progenitor mass range.

Combining the above expressions, the resulting energy density spectrum of the GWB from CCSNe becomes
\begin{align}
\Omega_{\text{CCSNe}}(f) =\; & \frac{8 \pi G}{3 c^2 H_0^3} f \int_0^{z_*} 
\frac{\dot\rho_*(z)}{(1+z)^{2}E(\Omega,z)} \notag \\
& \times A \exp\left[ - \frac{(f_s - f_{\text{peak}})^2}{2 \Delta^2} \right] \, \mathrm{d}z \int \Phi(m) \, \mathrm{d}m.
\label{eq:omega_CCSNe}
\end{align}
As noted in Ref.~\cite{Zhu:2010af}, the contribution of CCSNe to the overall event rate becomes negligible beyond redshift $z \sim 6$. We adopt a redshift cutoff of $z_* = 10$ in our analysis.

In addition, we include the CBC component of the GWB as a power law \cite{Zhu:2012xw},
\begin{equation}
\Omega_{\mathrm{CBC}}(f)=\Omega_{\mathrm{ref}}\!\left(\frac{f}{f_{\mathrm{ref}}}\right)^{\alpha}.
\end{equation}
Here $f_{\mathrm{ref}}=25\,\mathrm{Hz}$, and $\alpha=2/3$, which is appropriate for an inspiral-dominated CBC background spectrum; see Ref. \cite{Zhu:2012xw} for a detailed discussion. Although the CBC background is expected to deviate from a simple power-law model at frequencies above a few hundred hertz (due to merger and ringdown contributions), our inference on $\Omega_{\mathrm{ref}}$ is primarily informed by the low-frequency data where the detectors are most sensitive and where the power-law approximation is valid. Therefore, we consider the power-law model to be a good approximation for the CBC background in the full analysis band as specified in Sec. \ref{sec:III}.

\section{\label{sec:III}Upper limits on the GW energy emitted by CCSNe}

\begin{table*}[htbp]
    \centering
    \caption{\justifying
        Prior ranges and posterior estimates for the GWB parameters $\log_{10}\Omega_{\text{ref}}$, $\log_{10}A$, $f_{\text{peak}}$, and $\Delta$.
        The first column lists the parameters. The second column shows the uniform prior ranges for Prior I, whereas Prior II is the same as Prior I except that we extend the upper ends of both $f_{\text{peak}}$ and $\Delta$ to $\unit[2000]{Hz}$. The third and fourth columns present median estimates and the $95\%$ credible intervals for parameters of the CCSNe and CCSNe+CBC model, respectively. For posterior estimates, values outside and inside parentheses are obtained from Prior I and Prior II, respectively.
    }
    \label{Table:prior_posterior}
    \begin{ruledtabular}
    \begin{tabular}{lccc}
        Parameter & Prior I & Posterior for the CCSNe model & Posterior for the CCSNe+CBC model \\
        \hline
        $\log_{10}\Omega_{\text{ref}}$ & Uniform[$-13$, $-6$] & $\cdots$  & $-10.6^{+2.4}_{-2.3}$ ($-10.6^{+2.4}_{-2.3}$)\\
        $\log_{10}A$ & Uniform[$-14$, $-3$] & $-9.4^{+4.7}_{-4.3}$ ($-9.4^{+4.9}_{-4.4}$) & $-9.4^{+4.7}_{-4.3}$ ($-9.4^{+5.0}_{-4.4}$) \\
        $f_{\text{peak}}$ (Hz) & Uniform[20, 1000] & $520^{+457}_{-473}$ ($1044^{+919}_{-977}$) & $520^{+458}_{-475}$ ($1034^{+913}_{-961}$) \\
        $\Delta$ (Hz) & Uniform[20, 1000] & $502^{+474}_{-459}$ ($988^{+959}_{-928}$) & $497^{+478}_{-456}$ ($977^{+973}_{-914}$) \\
    \end{tabular}
    \end{ruledtabular}
\end{table*}


\begin{figure}[htbp]
    \centering
    \includegraphics[width=0.48\textwidth]{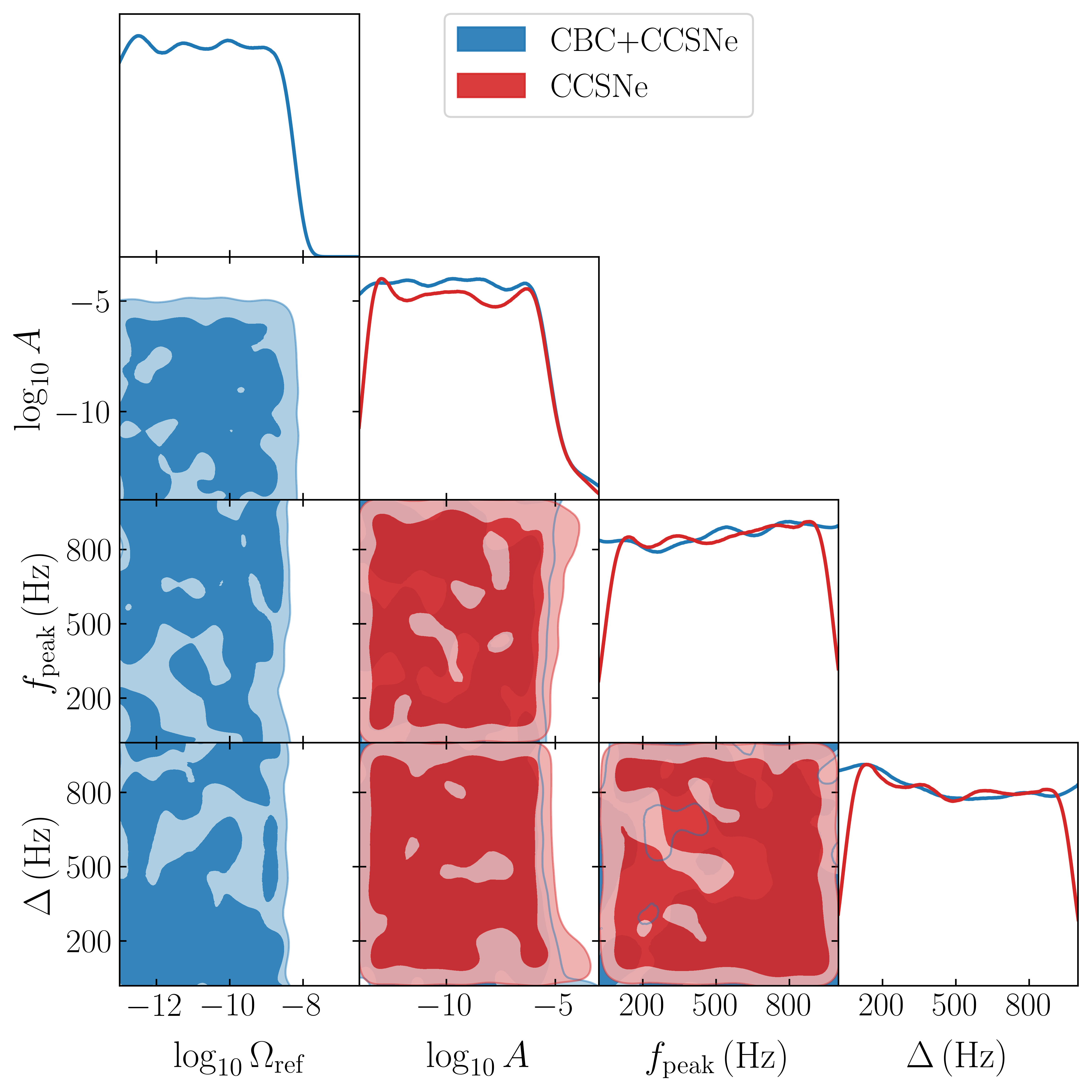}
    \caption{\justifying
        Posterior distributions of the GWB model parameters inferred from O3 data under Prior~I. The red contours correspond to the CCSNe model, while the blue contours represent the combined CBC+CCSNe model. The parameters shown are: reference energy density $\log_{10} \Omega_{\mathrm{ref}}$, amplitude $\log_{10} A$, peak frequency $f_{\mathrm{peak}}$, and bandwidth $\Delta$.
    }
    \label{fig:corner_plot}
\end{figure}

We analyze the O3 cross-correlation spectra \cite{LVK:O3opendata} from Advanced LIGO and Virgo. The data comprise coincident strain from the H1 (LIGO Hanford), L1 (LIGO Livingston), and V1 (Virgo) detectors, and we follow the methodology of Ref.~\cite{LVK-O3-SGWB}. Data segments are Hannwindowed, Fourier transformed, and inverse-noise weighted to construct the cross-correlation estimator; nonstationary intervals and instrumental lines are mitigated by time-domain gating and frequency notching. The frequency range of O3 binned spectra is $20$--$1726~\mathrm{Hz}$.
Assuming the cross-correlation estimator is approximately Gaussian, the likelihood function for model parameters $\boldsymbol{\Theta}$ reads

\begin{equation}
    p\big(\hat{C}_k | \mathbf{\Theta} \big) \propto \exp\left[-\frac{1}{2} \sum_k \left( \frac{\hat{C}_k - \Omega_{\text{M}}(f_k | \mathbf{\Theta})}{\sigma(f_k)} \right)^2 \right]\, ,
    \label{eq:likelihood}
\end{equation}
where $\hat{C}_k \equiv \hat{C}(f_k)$ is the measured cross-correlation estimator in the $k$th frequency bin and $\sigma(f_k)$ is its standard deviation. We consider two GWB models: (i) CCSNe, $\Omega_{\mathrm{M}}(f)=\Omega_{\mathrm{CCSNe}}(f)$ and (ii) CCSNe + CBC, $\Omega_{\mathrm{M}}(f)=\Omega_{\mathrm{CCSNe}}(f)+\Omega_{\mathrm{CBC}}(f)$. During sampling, $\Omega_{\mathrm{M}}(f_k|\Theta)$ is computed at each likelihood evaluation by numerically calculating Eq.~\eqref{eq:omega_CCSNe} on the analysis frequency grid $f_k$.

We explore the parameter space using nested sampling implemented via \texttt{Dynesty}~\cite{Speagle_2020} within the \texttt{Bilby} software package~\cite{Ashton:2018jfp}. To validate our analysis pipeline, we reproduced the results from Ref.~\cite{LVK-O3-SGWB} using the same dataset and computational setup. Our independent results show excellent agreement with FFig.~4 of Ref.~\cite{LVK-O3-SGWB}, confirming the effectiveness of our method (see the Appendix and Fig.~\ref{fig:pipeline_check}).
We note that the GWB analysis presented in Ref.~\cite{LVK-O3-SGWB} accounted for calibration uncertainties~\cite{Whelan:2012-calibration, Sun:2020wke-calibration, Romero:2021kby-calibration} for different detector baselines at the inference stage. The documentation of public O3 data products does not indicate whether or not calibration uncertainties are already included in the cross-correlation estimator and covariance. Since we successfully reproduce results from Ref.~\cite{LVK-O3-SGWB}, we do not explicitly account for calibration uncertainties in this work.

In Table~\ref{Table:prior_posterior}, we list the prior ranges for the model parameters.
For $\log_{10}\Omega_{\text{ref}}$ in the CCSNe+CBC model, we assume a uniform prior in logarithmic space over the range $[-13, -6]$, consistent with Ref.~\cite{LVK-O3-SGWB}.
For the peak frequency $f_{\text{peak}}$ and bandwidth $\Delta$ of the CCSNe spectrum, we assume a uniform prior between 20 and $\unit[1000]{Hz}$.
We also assume a uniform prior for the spectrum amplitude $\log_{10}A$, for which the prior range is set such that the GW energy\footnote{The GW energy is calculated by integrating the source-frame energy spectrum over the interval of $f_s\in[20,5000]\,\mathrm{Hz}$.} $E_{\text{GW}}$ is between $10^{-11}$ and $0.5 ~M_\odot c^2$ with 95\% prior probability. These priors, as listed in Table~\ref{Table:prior_posterior}, are denoted as Prior I. In order to account for peak emission above $\unit[1000]{Hz}$ and ultrabroadband spectrum, for Prior II we extend the upper ends of both $f_{\text{peak}}$ and $\Delta$ to $\unit[2000]{Hz}$, while keeping priors for the other two parameters unchanged.

\begin{figure}[htbp]
    \centering
    \includegraphics[width=0.48\textwidth]{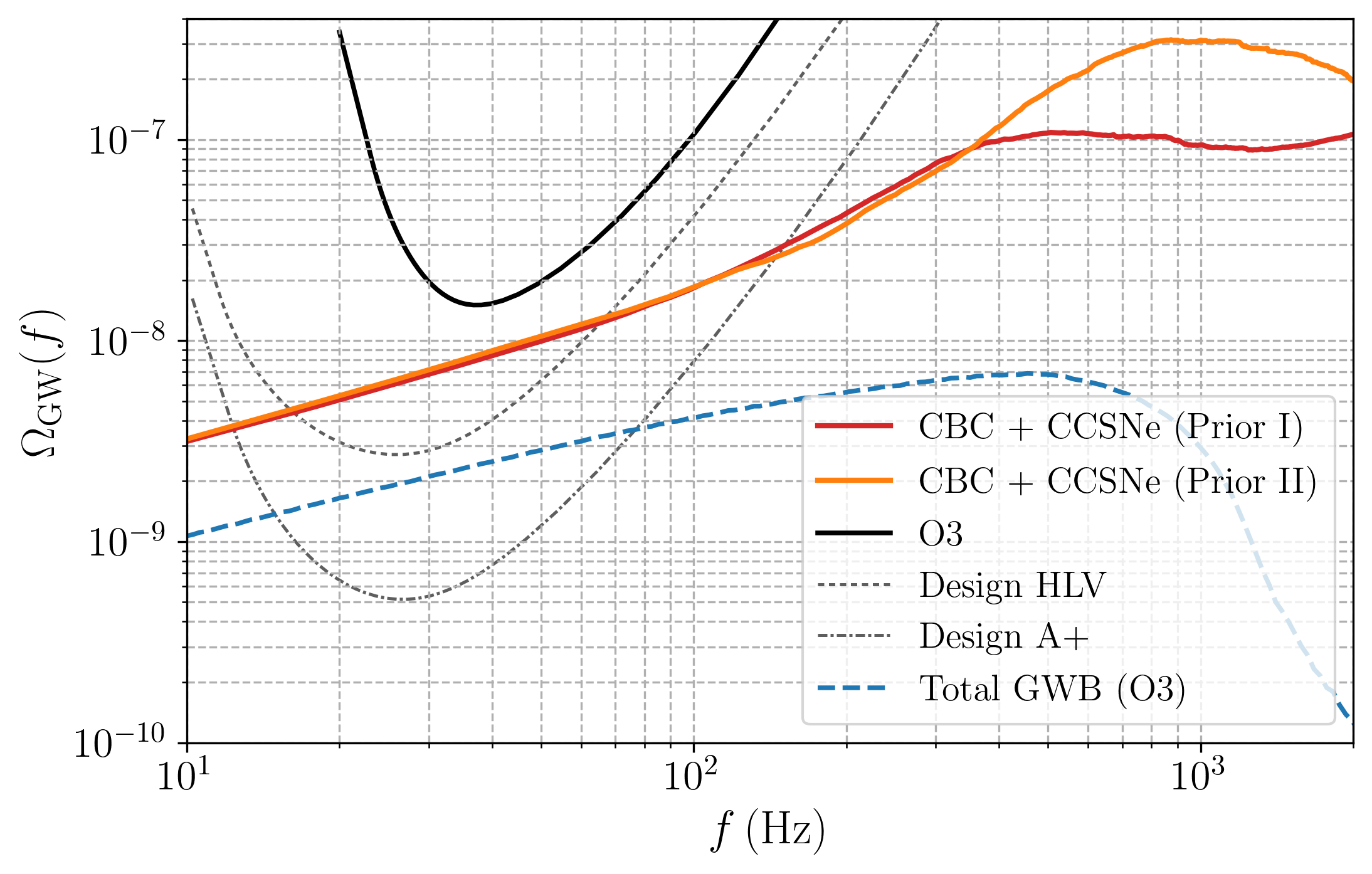}
    \caption{\justifying The 95\% credibility upper limit on $\Omega_{\rm GW}(f)$ for the CBC+CCSNe model under two different priors. Also shown are predictions for the GWB from CBCs informed by GWTC-3 (including contributions from binary neutron stars, binary black holes, and neutron star-black hole binaries), along with current and projected sensitivity curves~\cite{LVK-O3-SGWB}.}
    \label{fig:omega_gw}
\end{figure}

Figure.~\ref{fig:corner_plot} shows the posterior distributions of parameters for the CCSNe model (red) and the CCSNe+CBC model (blue). The posterior distributions are relatively broad and, for some parameters, remain close to the corresponding priors. In the nondetection regime, the main information gain is reflected in the exclusion of the high-amplitude region of the prior parameter space and is therefore best summarized through upper limits. In Table~\ref{Table:prior_posterior}, we also list the posterior estimates (including median values and 95\% credible intervals) of these model parameters for both sets of priors.
It can be seen that the inclusion of the CBC contribution in the GWB has a small impact on the CCSNe model posteriors. As an additional check, we perform a CBC-only inference using the same data, modeling the background as a pure power law with the inspiral index fixed to $\alpha=2/3$. We obtain $\log_{10}\Omega_{\mathrm{ref}} = -10.6^{+2.4}_{-2.2}$ ($95\%$ credible interval), which is statistically consistent with the posterior on $\Omega_{\mathrm{ref}}$ obtained in the CCSNe+CBC model. On the other hand, the CCSNe model posteriors are dependent on the prior ranges.
Figure.~\ref{fig:omega_gw} shows the 95\% credibility upper limit on $\Omega_{\rm GW}(f)$ based on the posterior predictive distribution derived from the posterior distributions of parameters for the CCSNe+CBC model. Specifically, we draw samples of the model parameters from the CCSNe+CBC posterior, compute the corresponding spectrum $\Omega_{\rm GW}(f)$ for each sample over the analysis band, and then construct the posterior predictive distribution of $\Omega_{\rm GW}(f)$ at each frequency bin. The plotted $95\%$ upper-limit curve is the pointwise 95th percentile of this distribution. It can be seen that our current limit is about a factor of 3 above the predicted CBC GWB signal amplitude as informed by the GW detection catalog.

Since we are most interested in the constraint on the GW energy, we convert the CCSNe model posteriors into a distribution of $E_{\text{GW}}$.
Under Prior I, the 95\% credibility upper limit on $E_{\text{GW}}$ is $1.1\times10^{-2}~M_\odot c^2$ and $9.9\times10^{-3}~M_\odot c^2$, for the CCSNe and CCSNe+CBC model, respectively.
Under Prior II, the upper limits on $E_{\text{GW}}$ is $3.3 \times 10^{-2}~M_\odot c^2$ (CCSNe) or $3.2 \times 10^{-2}~M_\odot c^2$ (CCSNe+CBC), a factor of 3 more conservative than Prior I.
In both cases, we can see that the inclusion of the CBC contribution leads to a modest ($\approx 10\%$) improvement in the constraint on $E_{\text{GW}}$.
Our results significantly improve upon earlier limit of $(0.49$–$1.98)\,M_\odot c^2$ obtained using data from initial LIGO~\cite{Zhu:2010af}.

Our upper limits are in the upper end of predictions of some phenomenological models~\cite{LIGOScientific:2016jvu}, which predict GW emission in the range of $10^{-4}$–$10^{-2}~M_\odot c^2$.
Comparing our constraints with the recent observational results from SN~2023ixf~\cite{LIGOScientific:2024jxh}, which set an upper limit of $\sim 10^{-4}~M_\odot c^2$ at 82 Hz, our results are weaker by about 2 orders of magnitude. However, their constraints apply at low frequencies. When extrapolated to the kilohertz regime, the energy bounds may increase to $10^{-2}$–$10^{-1}~M_\odot c^2$, consistent with our estimates.
It should be emphasized that the limit derived in this work is the \textit{average} GW energy emitted per CCSN event across the cosmic population, which is complementary to single-event emission of a specific nearby supernova.

\begin{figure*}[htbp]
    \centering
        \includegraphics[width=\linewidth]{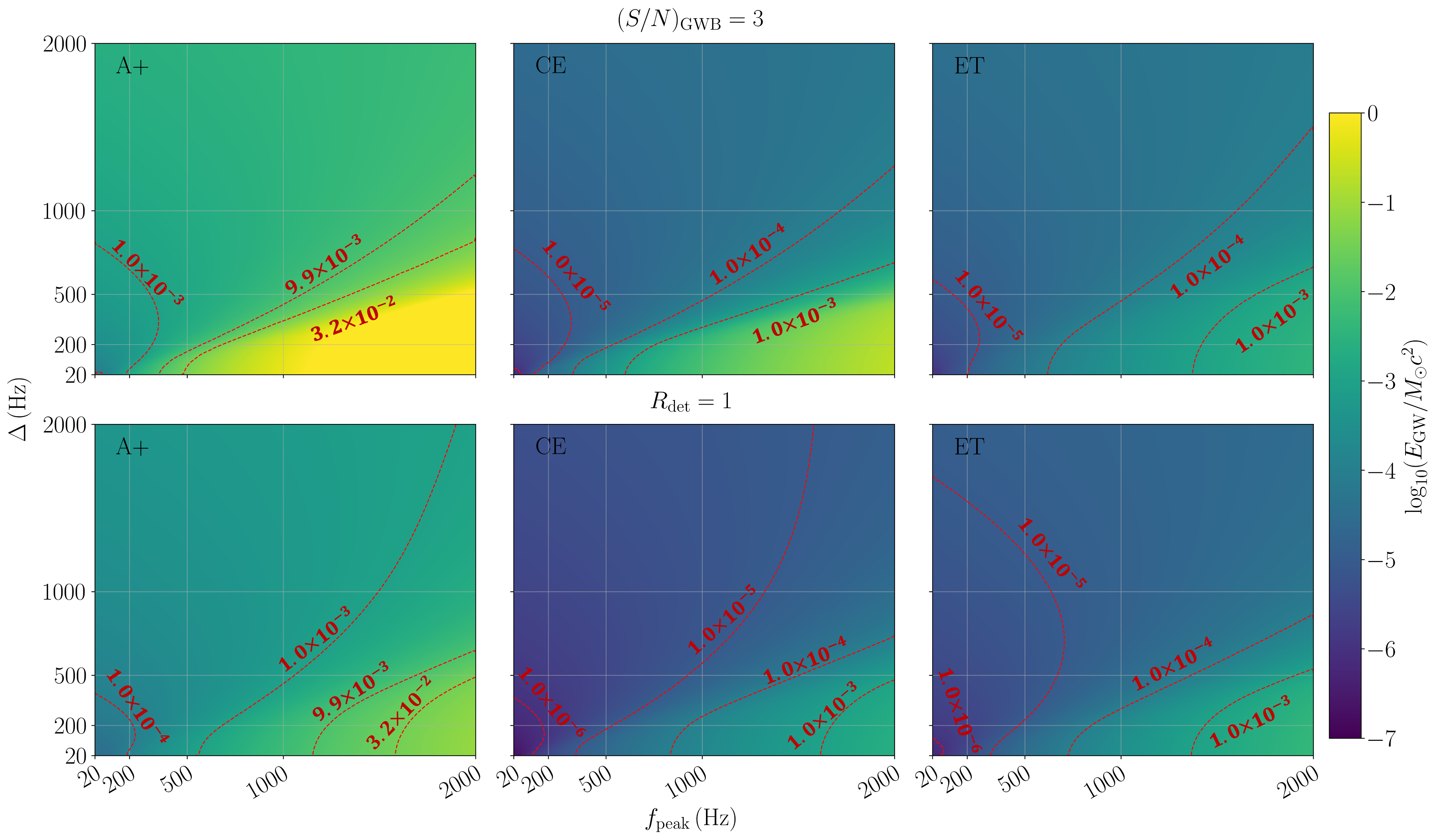}
    \caption{\justifying
        Assuming a generic Gaussian emission spectrum, the required GW energy $E_{\mathrm{GW}}$ as a function of peak frequency $f_{\text{peak}}$ and spectral width $\Delta$ to achieve $(S/N)_{\mathrm{GWB}} = 3$ (top row) and a detection rate ($R_{\rm det}$) of one CCSNe event per year (bottom row).
        Panels correspond to different detectors A+ (left), CE (middle), and ET (right). Red dashed lines indicate constant levels of $E_{\mathrm{GW}}$.
    }
    \label{fig:f0_delta_E}
\end{figure*}

\section{\label{sec:IV}Detection prospects}

\begin{figure*}[htbp]
    \centering
        \includegraphics[width=\linewidth]{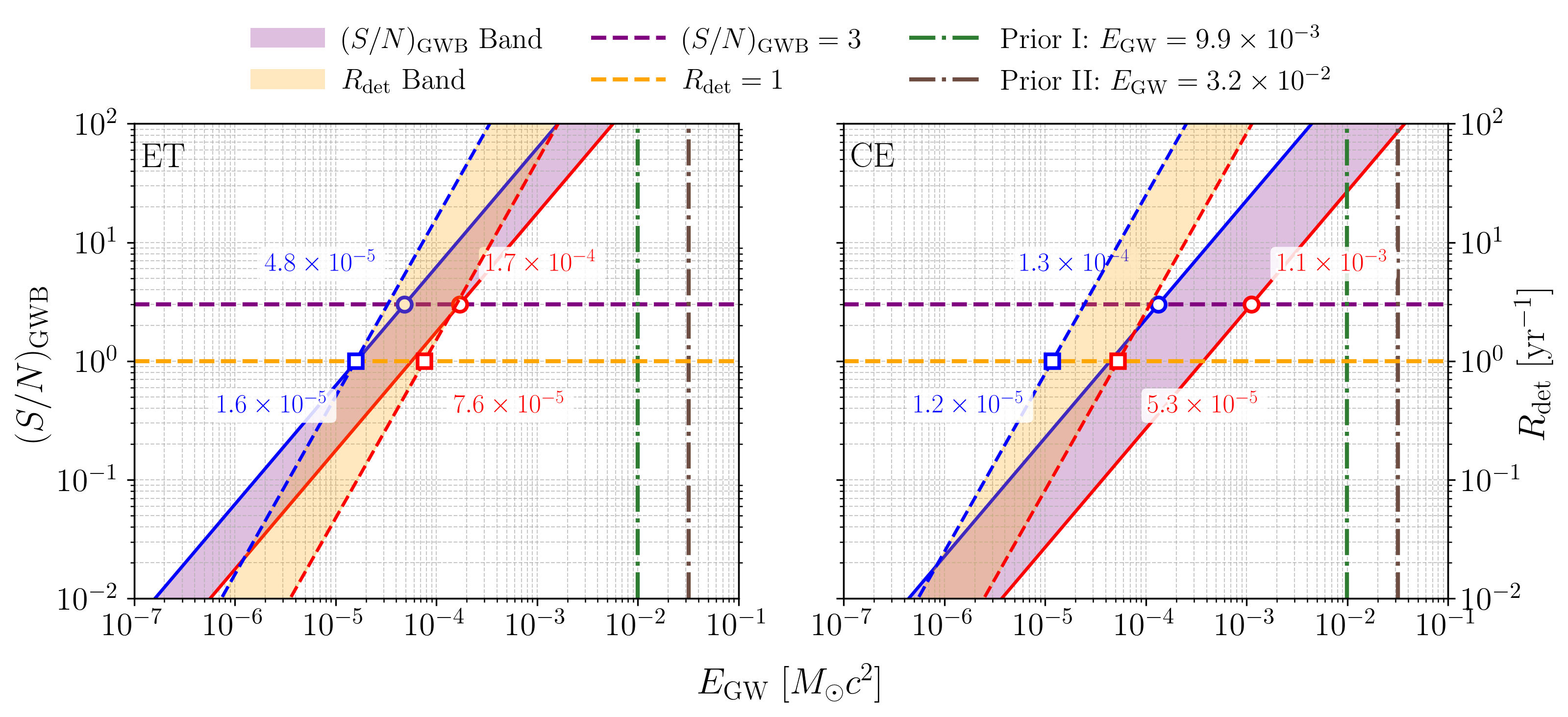}
    \caption{\justifying The signal to noise ratio of the GWB $(S/N)_{\rm GWB}$ (solid lines with shaded band) or the detection rate $R_{\rm det}$ (dashed lines with shaded band) of individual CCSNe events as a function the GW energy emitted by CCSNe events for the ET (left) or CE (right) detector. Blue and red curves denote two CCSNe spectral models (see text for details): Ott+11 s27-fheat1.05 (blue) and And+19 m15r (red). The crossing between the $(S/N)_{\rm GWB}$ lines and the $R_{\rm det}$ lines with their respective thresholds (horizontal lines) are marked with circles and squares (with the GW energy values listed), respectively. Two vertical lines are the upper limits derived in this work.}
    \label{fig:detection_SNR_Rate}
\end{figure*}

In this section, we explore the detection prospects of both the GWB and individual CCSNe events for future detectors.
Assuming Gaussian detector noise, the optimal signal-to-noise ratio ($S/N$) for a cross-correlation analysis over an observation time $T$ is given by~\cite{Allen:1997ad}
\begin{equation}
    (S/N)_{\mathrm{GWB}}^2 = \frac{9 H_0^4}{50 \pi^4} T \int_0^{\infty} \mathrm{d}f \, \frac{\gamma^2(f) \, \Omega_{\mathrm{GW}}^2(f)}{f^6 P_1(f) P_2(f)},
\label{eq:snr_sgwb}
\end{equation}
where $P_1(f)$ and $P_2(f)$ denote the power spectral density of the two detectors, and $\gamma(f)$ is the overlap reduction function (ORF). In this work, we adopt a detection threshold of $(S/N)_{\mathrm{GWB}} = 3$, accumulated over one year of observation, to assess the detectability of the GWB from CCSNe.

We further evaluate the detectability of individual CCSNe events using the matched-filter method, assuming optimal detector alignment and isotropic GW emission. The matched-filter $S/N$ can be expressed as~\cite{Flanagan:1997sx, Andresen2019}
\begin{equation}
(S/N)_{\mathrm{CCSNe}}^2 = 4 \int_0^{\infty} \mathrm{d}f \, \frac{|\tilde{h}(f)|^2}{S(f)} = \int_0^{\infty} \mathrm{d}f \, \frac{h_c^2}{f^2 S(f)},
\label{eq:snr_ccSNe}
\end{equation}
where the characteristic strain is defined as
\begin{equation}
    h_c(f) = \sqrt{\frac{2G}{\pi^2 c^3 D_{L}^2(z)}\frac{\mathrm{d}E}{\mathrm{d}f_s}}.
\end{equation}
To calculate the detectable event rate ($R_\text{det}$), we follow the approach in Ref.~\cite{Szczepanczyk:2021bka}, adopting a matched-filter detection threshold of $(S/N)_{\mathrm{CCSNe}} = 20$ and assuming a detection efficiency of $50\%$.
We consider three detector configurations: A+~\cite{LIGO:2018instrument}, the Einstein Telescope (ET)~\cite{Hild:2010id}, and the Cosmic Explorer (CE)~\cite{Evans:2021gyd}. Specifically, we consider the ET-D configuration with three nearly colocated detectors, and a pair of two detectors with $40$ and $20~\rm{km}$ arm lengths for CE. For simplicity, we refer to these respective detector pairs as ET and CE throughout this work. The sensitivity curves for these detectors are obtained from Ref. \cite{LIGOScientific:2016wof}. Details of the detector configurations and the corresponding ORFs can be found in Ref.~\cite{Gupta2024}. The ORFs used in our analysis are computed following Ref.~\cite{Allen:2001ay}.

Figure~\ref{fig:f0_delta_E} illustrates the detection prospects for the GWB (top panels) or individual CCSNe events (bottom panels) for three future detectors: A+ (left), CE (middle), and ET (right), assuming Gaussian emission spectrum as described by Eq. (\ref{eq:energy_spectrum}).
Each panel is color coded by $\log_{10}(E_{\mathrm{GW}}/M_\odot c^2)$, the required GW energy emitted a single CCSNe event, as a function of the peak frequency $f_{\mathrm{peak}}$ and the spectral width $\Delta$, to achieve a GWB detection with $(S/N)_{\mathrm{GWB}} = 3$ or a detection rate of one CCSNe event per year ($R_{\mathrm{det}} = 1$).

The GW energy required for detection spans a wide range—typically over 2-3 orders of magnitude—depending on the choice of spectral parameters and detector sensitivity. Notably, the detectability is strongly dependent on both $f_{\mathrm{peak}}$ and $\Delta$. We find that high-frequency, narrow-band signals (e.g., $f_{\mathrm{peak}} \gtrsim 1000$, $\Delta \lesssim 500~\mathrm{Hz}$) require significantly more GW energy to be detectable. This is due to the increasing detector noise at high frequencies and, in the case of GWB detection, the reduced cross-correlation sensitivity defined by the overlap reduction function. In comparison, low-frequency signals ($f_{\mathrm{peak}} \lesssim 200~\mathrm{Hz}$) allow for detection with GW energy $E_{\mathrm{GW}}$ lower by a factor of 10-100.
Across three detectors, we can see that the required GW energy for the detection of the GWB is generally higher than that for a detection rate of one single CCSNe event per year.
This indicates that individual CCSNe events are likely to be detected prior to their cumulative GWB becoming observable.

Each panel of Fig.~\ref{fig:f0_delta_E} includes several red dashed contours, indicating constant levels of $E_{\mathrm{GW}}$. For the A+ panels, we include two lines corresponding to upper limits derived in this work for Prior I ($E_{\mathrm{GW}}=9.9\times10^{-3}~M_\odot c^2$) and Prior II ($E_{\mathrm{GW}}=3.2\times10^{-2}~M_\odot c^2$). 
As a guideline, we can see that the A+ detector would be able to further improve upper limits on $E_{\mathrm{GW}}$ to $10^{-3}$ and $10^{-4}~M_\odot c^2$ levels for low-frequency signals ($f_{\mathrm{peak}} \lesssim 200~\mathrm{Hz}$), through the GWB and single-event searches, respectively.
For third-generation detectors such as ET and CE, the detection prospects are more promising.
There are chances for detection of the GWB if the GW energy is above $10^{-4}~M_\odot c^2$ for a large fraction of the parameter space.
For single events, the detection limit can be pushed to $10^{-5}~M_\odot c^2$ or even $10^{-6}~M_\odot c^2$ for low-frequency signals.

So far our calculations are based on the assumption that the GW emission spectrum from CCSNe events can be approximated by a Gaussian function.
To illustrate the impact of CCSN waveform modeling uncertainties on our detectability forecasts, we consider two CCSNe spectral models from Fig.~3 of Ref.~\cite{Szczepanczyk:2021bka}: the Ott+11 s27-fheat1.05 model (blue) and the And+19 m15r model (red).\footnote{A comprehensive survey of all available CCSN numerical-simulation waveforms is impractical. We therefore evaluated six waveforms shown in Fig.~3 of Ref.~\cite{Szczepanczyk:2021bka} and selected these two representative models with intermediate (i.e., nonextreme) results for demonstration purposes.} We keep the spectral shape of these two models and rescale the amplitude\footnote{This is for illustrative purpose only as we aim to find a plausible range of required GW energy for detection of GWs from CCSNe with third-generation detectors.} to calculate the GWB single-to-noise ratio $(S/N)_{\mathrm{GWB}}$ and single-event detection rate $R_\text{det}$ for ET and CE.
Results are shown in Fig. \ref{fig:detection_SNR_Rate}.
Solid lines indicate the $(S/N)_{\mathrm{GWB}}$ as a function of the average GW energy per CCSNe, while dashed lines show the corresponding detectable event rate per year. The purple and orange shaded bands represent model-dependent uncertainties. In the same figure, the gray and gold dashed horizontal lines mark the detection thresholds $(S/N)_{\mathrm{GWB}} = 3$ and $R_\text{det} = 1$, respectively. The vertical lines correspond to the GW energy constraints obtained in this work.

In Fig.~\ref{fig:detection_SNR_Rate}, one can draw consistent conclusions as in Fig.~\ref{fig:f0_delta_E}.
First, the requirement on the GW energy for a detection of the GWB is generally higher than that for individual CCSNe events.
This is most significant for CE, with two differing by a factor of 10, whereas the difference is around $3$ for ET.
This can be attributed to differences in detector noise characteristics and the ORF.
The GWB can be detectable with third-generation detectors if the GW energy is $\gtrsim 10^{-4}~M_\odot c^2$, whereas a detection of single CCSNe event is possible if 
$E_{\text{GW}} \gtrsim 10^{-5}~M_\odot c^2$.

\section{\label{sec:V}Conclusions}

Core-collapse supernovae have long been considered to be an important source of GWs. While direct detections from CCSNe remain elusive, constraining their average GW energy emission is crucial for evaluating the detectability of CCSNe signals. In this work, we use cross-correlation measurements from the third observing run of LIGO and Virgo detectors to place upper limits on the average GW energy emitted per CCSNe event\footnote{At the time of writing this paper, results of a GWB search using the first part of O4 data were released \cite{LVK-O4a-sgwb}. We expect that our upper limits would be slightly improved if O4a data were used.}.
By adopting a generic Gaussian spectrum model, we drive constraints that are marginalized over a wide range of emission peak frequency and spectral width. Furthermore, we assess the detection prospects with third-generation ground-based detectors,including ET and CE, for both individual events and the GWB produced by a cosmic population of CCSNe sources.

Our Bayesian analysis yields a $95\%$ credibility upper limit on the average GW energy per CCSNe, improving upon previous bounds of $(0.49$–$1.98)~M_\odot c^2$ from Ref.~\cite{Zhu:2010af} by approximately 2 orders of magnitude. Specifically, we obtain upper bounds of $1.1\times10^{-2} ~M_\odot c^2$ while considering only contributions from CCSNe to the GWB, and $9.9\times10^{-3} ~M_\odot c^2$ when CBCs are included.
Our upper limits are in the upper end of the most optimistic theoretical predictions, e.g., long-lasting bar-mode emission mechanisms~\cite{LIGOScientific:2016jvu}.
As detector sensitivities continue to improve, we anticipate that upcoming observations with advanced detectors will place astrophysically interesting bounds on the GW emission from CCSNe.

Finally, we compare the detectability of the GWB and individual CCSNe events for future detectors. In general, individual events are expected to be detectable prior to the GWB.
For third-generation detectors such as ET and CE, we find that the GWB can be detectable if the GW energy is $\gtrsim 10^{-4}~M_\odot c^2$, and a detection of a single CCSNe event is possible within one year of operation if 
$E_{\text{GW}} \gtrsim 10^{-5}~M_\odot c^2$.
We also find that for GW emission with a peak frequency above $\unit[1000]{Hz}$, the sensitivity of ET or CE to both the GWB and single events decreases by a factor of $\sim 10$ in comparison to signals with peak frequencies below $\unit[500]{Hz}$. This highlights the scientific potential of dedicated kilohertz detectors \cite{Miao2018,Martynov2019,Nemo20PASA,ZhangTeng23PRX}.

\begin{acknowledgments}
We thank the anonymous referees for useful comments on
the manuscript. This work is supported by the National Key Research and Development Program of China (No. 2023YFC2206704), the Fundamental Research Funds for the Central Universities, and the Supplemental Funds for Major Scientific Research Projects of Beijing Normal University (Zhuhai) under Project ZHPT2025001. This research has made use of data or software obtained from the Gravitational Wave Open Science Center\cite{gwosc}, a service of the LIGO Scientific Collaboration, the Virgo Collaboration, and KAGRA. This material is based upon work supported by NSF's LIGO Laboratory which is a major facility fully funded by the National Science Foundation, as well as the Science and Technology Facilities Council (STFC) of the United Kingdom, the Max-Planck-Society (MPS), and the State of Niedersachsen/Germany for support of the construction of Advanced LIGO and construction and operation of the GEO600 detector. Additional support for Advanced LIGO was provided by the Australian Research Council. Virgo is funded, through the European Gravitational Observatory (EGO), by the French Centre National de Recherche Scientifique (CNRS), the Italian Istituto Nazionale di Fisica Nucleare (INFN) and the Dutch Nikhef, with contributions by institutions from Belgium, Germany, Greece, Hungary, Ireland, Japan, Monaco, Poland, Portugal, Spain. KAGRA is supported by Ministry of Education, Culture, Sports, Science and Technology (MEXT), Japan Society for the Promotion of Science (JSPS) in Japan; National Research Foundation (NRF) and Ministry of Science and ICT (MSIT) in Korea; Academia Sinica (AS) and National Science and Technology Council (NSTC) in Taiwan.
\end{acknowledgments}

\section{Data availability}
The data that support the findings of this article are openly available\cite{data-available}
\appendix

\section{\label{sec:appendix}Reproduction of LIGO-VIRGO-KAGRA (LVK) results}

To validate our Bayesian inference method, we reproduce the LVK O3 isotropic stochastic background analysis presented in Ref.~\cite{LVK-O3-SGWB}. Specifically, we analyze the same publicly released O3 cross-correlation data products and adopt the same two-parameter power-law model, $\Omega_{\rm GW}(f)=\Omega_{\rm ref}(f/f_{\rm ref})^{\alpha}$, with the same reference frequency and prior choices as in Ref.~\cite{LVK-O3-SGWB}. 

\begin{figure}[htbp]
    \centering
    \includegraphics[width=0.48\textwidth]{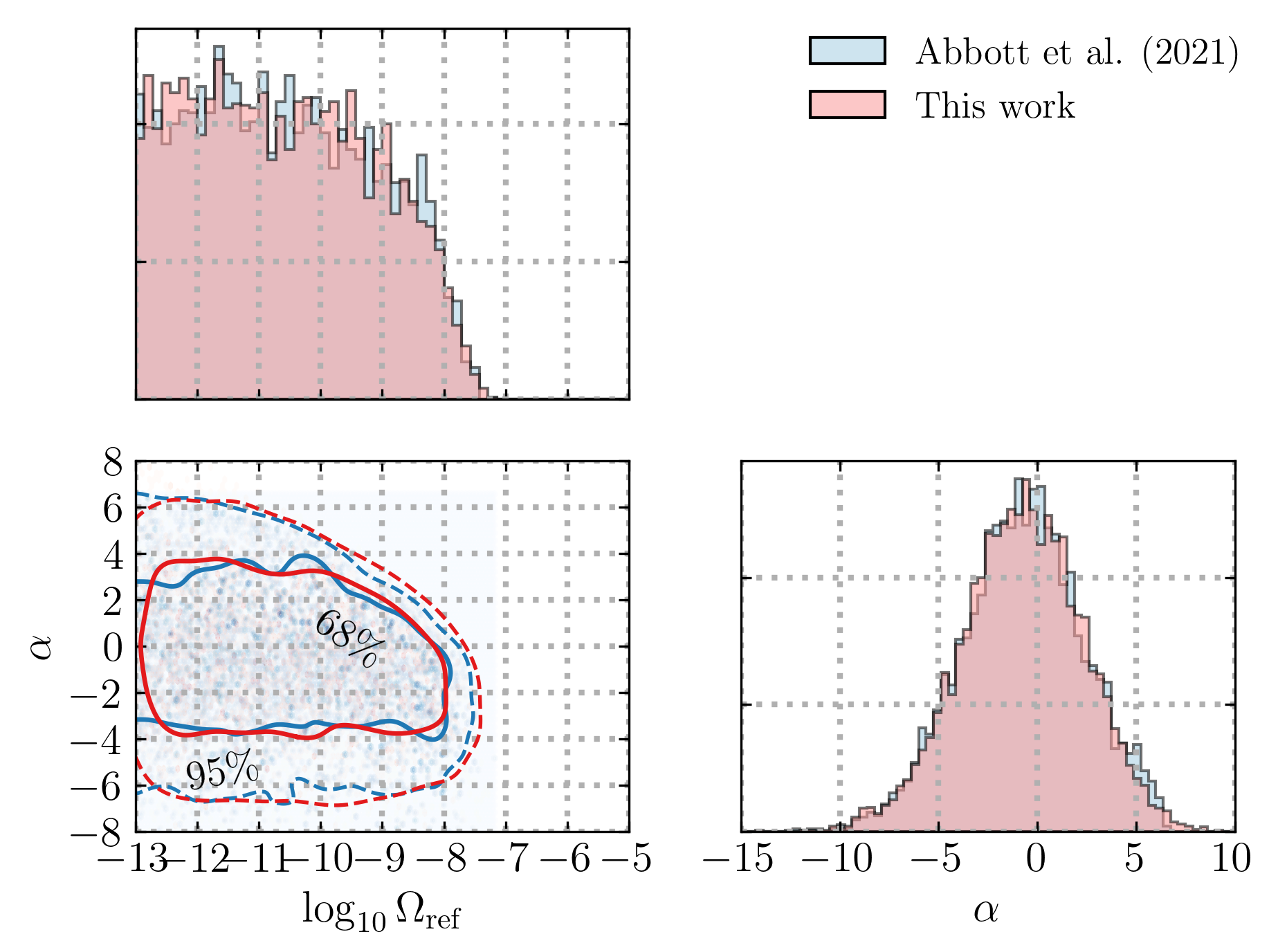}
    \caption{\justifying
Posterior distributions of the power-law GWB model parameters $(\log_{10}\Omega_{\rm ref},\,\alpha)$ using the O3 cross-correlation data. The blue contours correspond to the results reported in Ref.~\cite{LVK-O3-SGWB}, while the red contours show the posteriors obtained in this work using the same data products. The diagonal panels show the marginalized one-dimensional posteriors, and the lower-left panel shows the joint posterior. Solid (dashed) contours enclose the 68\% (95\%) credible regions.}
    \label{fig:pipeline_check}
\end{figure}

Figure~\ref{fig:pipeline_check} compares the resulting posterior distributions with those reported in Ref.~\cite{LVK-O3-SGWB}. We find excellent agreement in both the one-dimensional marginals and the two-dimensional credible regions, with any small differences consistent with finite-sampling uncertainty (and, potentially, minor implementation details). This reproduction test provides an independent validation of our analysis framework before applying it to the CCSNe and CCSNe+CBC models considered in this work.

\bibliography{CCSNe}

@article{LIGOScientific:2016aoc,
    author = "Abbott, B. P. and others",
    collaboration = "LIGO, Virgo",
    title = "{Observation of Gravitational Waves from a Binary Black Hole Merger}",
    eprint = "1602.03837",
    archivePrefix = "arXiv",
    primaryClass = "gr-qc",
    reportNumber = "LIGO-P150914",
    doi = "10.1103/PhysRevLett.116.061102",
    journal = "Phys. Rev. Lett.",
    volume = "116",
    number = "6",
    pages = "061102",
    year = "2016"
}

@article{VIRGO:2014yos,
    author = "Acernese, F. and others",
    collaboration = "Virgo",
    title = "{Advanced Virgo: a second-generation interferometric gravitational wave detector}",
    eprint = "1408.3978",
    archivePrefix = "arXiv",
    primaryClass = "gr-qc",
    doi = "10.1088/0264-9381/32/2/024001",
    journal = "Class. Quant. Grav.",
    volume = "32",
    number = "2",
    pages = "024001",
    year = "2015"
}

@article{LIGOScientific:2014pky,
    author = "Aasi, J. and others",
    collaboration = "LIGO",
    title = "{Advanced LIGO}",
    eprint = "1411.4547",
    archivePrefix = "arXiv",
    primaryClass = "gr-qc",
    doi = "10.1088/0264-9381/32/7/074001",
    journal = "Class. Quant. Grav.",
    volume = "32",
    pages = "074001",
    year = "2015"
}

@article{KAGRA:2018plz,
    author = "Akutsu, T. and others",
    collaboration = "KAGRA",
    title = "{KAGRA: 2.5 Generation Interferometric Gravitational Wave Detector}",
    eprint = "1811.08079",
    archivePrefix = "arXiv",
    primaryClass = "gr-qc",
    reportNumber = "JGW-P1809243",
    doi = "10.1038/s41550-018-0658-y",
    journal = "Nature Astron.",
    volume = "3",
    number = "1",
    pages = "35--40",
    year = "2019"
}

@article{Zhu:2011bd,
    author = "Zhu, Xing-Jiang and Howell, E. and Regimbau, T. and Blair, D. and Zhu, Zong-Hong",
    title = "{Stochastic Gravitational Wave Background from Coalescing Binary Black Holes}",
    eprint = "1104.3565",
    archivePrefix = "arXiv",
    primaryClass = "gr-qc",
    reportNumber = "LIGO-P1000181-V3",
    doi = "10.1088/0004-637X/739/2/86",
    journal = "Astrophys. J.",
    volume = "739",
    pages = "86",
    year = "2011"
}

@article{Marassi:2011si,
    author = "Marassi, S. and Schneider, R. and Corvino, G. and Ferrari, V. and Portegies Zwart, S.",
    title = "{Imprint of the merger and ring-down on the gravitational wave background from black hole binaries coalescence}",
    eprint = "1111.6125",
    archivePrefix = "arXiv",
    primaryClass = "astro-ph.CO",
    doi = "10.1103/PhysRevD.84.124037",
    journal = "Phys. Rev. D",
    volume = "84",
    pages = "124037",
    year = "2011"
}

@article{Wu:2011ac,
    author = "Wu, C. and Mandic, V. and Regimbau, T.",
    title = "{Accessibility of the Gravitational-Wave Background due to Binary Coalescences to Second and Third Generation Gravitational-Wave Detectors}",
    eprint = "1112.1898",
    archivePrefix = "arXiv",
    primaryClass = "gr-qc",
    doi = "10.1103/PhysRevD.85.104024",
    journal = "Phys. Rev. D",
    volume = "85",
    pages = "104024",
    year = "2012"
}

@article{Zhu:2012xw,
    author = "Zhu, Xing-Jiang and Howell, Eric J. and Blair, David G. and Zhu, Zong-Hong",
    title = "{On the gravitational wave background from compact binary coalescences in the band of ground-based interferometers}",
    eprint = "1209.0595",
    archivePrefix = "arXiv",
    primaryClass = "gr-qc",
    reportNumber = "LIGO-P1200105-V2",
    doi = "10.1093/mnras/stt207",
    journal = "Mon. Not. Roy. Astron. Soc.",
    volume = "431",
    number = "1",
    pages = "882--899",
    year = "2013"
}

@ARTICLE{Miao2018,
       author = {{Miao}, Haixing and {Yang}, Huan and {Martynov}, Denis},
        title = "{Towards the design of gravitational-wave detectors for probing neutron-star physics}",
      journal = {\prd},
         year = 2018,
        month = aug,
       volume = {98},
       number = {4},
          eid = {044044},
        pages = {044044},
          doi = {10.1103/PhysRevD.98.044044},
archivePrefix = {arXiv},
       eprint = {1712.07345},
 primaryClass = {gr-qc},
       adsurl = {https://ui.adsabs.harvard.edu/abs/2018PhRvD..98d4044M}
}

@ARTICLE{Nemo20PASA,
       author = {{Ackley}, K. and others},
        title = "{Neutron Star Extreme Matter Observatory: A kilohertz-band gravitational-wave detector in the global network}",
      journal = {Publications of the Astronomical Society of Australia},
         year = 2020,
        month = nov,
       volume = {37},
          eid = {e047},
        pages = {e047},
          doi = {10.1017/pasa.2020.39},
archivePrefix = {arXiv},
       eprint = {2007.03128},
 primaryClass = {astro-ph.HE},
       adsurl = {https://ui.adsabs.harvard.edu/abs/2020PASA...37...47A}
}

@ARTICLE{ZhangTeng23PRX,
       author = {{Zhang}, Teng and {Yang}, Huan and {Martynov}, Denis and {Schmidt}, Patricia and {Miao}, Haixing},
        title = "{Gravitational-Wave Detector for Postmerger Neutron Stars: Beyond the Quantum Loss Limit of the Fabry-Perot-Michelson Interferometer}",
      journal = {Physical Review X},
         year = 2023,
        month = apr,
       volume = {13},
       number = {2},
          eid = {021019},
        pages = {021019},
          doi = {10.1103/PhysRevX.13.021019},
archivePrefix = {arXiv},
       eprint = {2212.12144},
 primaryClass = {gr-qc},
       adsurl = {https://ui.adsabs.harvard.edu/abs/2023PhRvX..13b1019Z}
}

@article{Buonanno:2004tp,
    author = "Buonanno, Alessandra and Sigl, Gunter and Raffelt, Georg G. and Janka, Hans-Thomas and Muller, Ewald",
    title = "{Stochastic gravitational wave background from cosmological supernovae}",
    eprint = "astro-ph/0412277",
    archivePrefix = "arXiv",
    doi = "10.1103/PhysRevD.72.084001",
    journal = "Phys. Rev. D",
    volume = "72",
    pages = "084001",
    year = "2005"
}

@article{Sandick:2006sm,
    author = "Sandick, Pearl and Olive, Keith A. and Daigne, Frederic and Vangioni, Elisabeth",
    title = "{Gravitational Waves from the First Stars}",
    eprint = "astro-ph/0603544",
    archivePrefix = "arXiv",
    reportNumber = "UMN-TH-2435-06, FTPI-MINN-06-07",
    doi = "10.1103/PhysRevD.73.104024",
    journal = "Phys. Rev. D",
    volume = "73",
    pages = "104024",
    year = "2006"
}

@article{Zhu:2010af,
    author = "Zhu, Xing-Jiang and Howell, Eric and Blair, David",
    title = "{Observational upper limits on the gravitational wave production of core collapse supernovae}",
    eprint = "1008.0472",
    archivePrefix = "arXiv",
    primaryClass = "gr-qc",
    reportNumber = "LIGO-P1000076-V2",
    doi = "10.1111/j.1745-3933.2010.00965.x",
    journal = "Mon. Not. Roy. Astron. Soc.",
    volume = "409",
    pages = "L132--L136",
    year = "2010"
}

@article{Smartt:2009zr,
    author = "Smartt, Stephen J.",
    title = "{Progenitors of core-collapse supernovae}",
    eprint = "0908.0700",
    archivePrefix = "arXiv",
    primaryClass = "astro-ph.SR",
    doi = "10.1146/annurev-astro-082708-101737",
    journal = "Ann. Rev. Astron. Astrophys.",
    volume = "47",
    pages = "63--106",
    year = "2009"
}

@article{Ferrari:1998jf,
    author = "Ferrari, Valeria and Matarrese, Sabino and Schneider, Raffaella",
    title = "{Stochastic background of gravitational waves generated by a cosmological population of young, rapidly rotating neutron stars}",
    eprint = "astro-ph/9806357",
    archivePrefix = "arXiv",
    doi = "10.1046/j.1365-8711.1999.02207.x",
    journal = "Mon. Not. Roy. Astron. Soc.",
    volume = "303",
    pages = "258",
    year = "1999"
}

@article{Regimbau:2001kx,
    author = "Regimbau, T. and de Freitas Pacheco, Jose A.",
    title = "{Cosmic background of gravitational waves from rotating neutron stars}",
    eprint = "astro-ph/0105260",
    archivePrefix = "arXiv",
    doi = "10.1051/0004-6361:20011005",
    journal = "Astron. Astrophys.",
    volume = "376",
    pages = "381",
    year = "2001"
}

@article{Zhu:2011pt,
    author = "Zhu, Xing-Jiang and Fan, Xi-Long and Zhu, Zong-Hong",
    title = "{Stochastic Gravitational Wave Background from Neutron Star r-mode Instability Revisited}",
    eprint = "1102.2786",
    archivePrefix = "arXiv",
    primaryClass = "astro-ph.CO",
    doi = "10.1088/0004-637X/729/1/59",
    journal = "Astrophys. J.",
    volume = "729",
    pages = "59",
    year = "2011"
}

@article{Lasky:2013jfa,
    author = "Lasky, Paul D. and Bennett, Mark F. and Melatos, Andrew",
    title = "{Stochastic gravitational wave background from hydrodynamic turbulence in differentially rotating neutron stars}",
    eprint = "1302.6033",
    archivePrefix = "arXiv",
    primaryClass = "astro-ph.HE",
    doi = "10.1103/PhysRevD.87.063004",
    journal = "Phys. Rev. D",
    volume = "87",
    number = "6",
    pages = "063004",
    year = "2013"
}

@article{Kibble:1976sj,
    author = "Kibble, T. W. B.",
    title = "{Topology of Cosmic Domains and Strings}",
    reportNumber = "ICTP/75/5",
    doi = "10.1088/0305-4470/9/8/029",
    journal = "J. Phys. A",
    volume = "9",
    pages = "1387--1398",
    year = "1976"
}

@article{Sarangi:2002yt,
    author = "Sarangi, Saswat and Tye, S. H. Henry",
    title = "{Cosmic string production towards the end of brane inflation}",
    eprint = "hep-th/0204074",
    archivePrefix = "arXiv",
    reportNumber = "CLNS-02-1786",
    doi = "10.1016/S0370-2693(02)01824-5",
    journal = "Phys. Lett. B",
    volume = "536",
    pages = "185--192",
    year = "2002"
}

@article{Damour:2004kw,
    author = "Damour, Thibault and Vilenkin, Alexander",
    title = "{Gravitational radiation from cosmic (super)strings: Bursts, stochastic background, and observational windows}",
    eprint = "hep-th/0410222",
    archivePrefix = "arXiv",
    doi = "10.1103/PhysRevD.71.063510",
    journal = "Phys. Rev. D",
    volume = "71",
    pages = "063510",
    year = "2005"
}

@article{LIGOScientific:2017ikf,
    author = "Abbott, B. P. and others",
    collaboration = "LIGO, Virgo",
    title = "{Constraints on cosmic strings using data from the first Advanced LIGO observing run}",
    eprint = "1712.01168",
    archivePrefix = "arXiv",
    primaryClass = "gr-qc",
    doi = "10.1103/PhysRevD.97.102002",
    journal = "Phys. Rev. D",
    volume = "97",
    number = "10",
    pages = "102002",
    year = "2018"
}

@article{Lopez:2013mqa,
    author = "Lopez, Alejandro and Freese, Katherine",
    title = "{First Test of High Frequency Gravity Waves from Inflation using Advanced LIGO}",
    eprint = "1305.5855",
    archivePrefix = "arXiv",
    primaryClass = "astro-ph.HE",
    doi = "10.1088/1475-7516/2015/01/037",
    journal = "JCAP",
    volume = "01",
    pages = "037",
    year = "2015"
}

@article{Dev:2016feu,
    author = "Dev, P. S. Bhupal and Mazumdar, A.",
    title = "{Probing the Scale of New Physics by Advanced LIGO/VIRGO}",
    eprint = "1602.04203",
    archivePrefix = "arXiv",
    primaryClass = "hep-ph",
    doi = "10.1103/PhysRevD.93.104001",
    journal = "Phys. Rev. D",
    volume = "93",
    number = "10",
    pages = "104001",
    year = "2016"
}

@article{Marzola:2017jzl,
    author = "Marzola, Luca and Racioppi, Antonio and Vaskonen, Ville",
    title = "{Phase transition and gravitational wave phenomenology of scalar conformal extensions of the Standard Model}",
    eprint = "1704.01034",
    archivePrefix = "arXiv",
    primaryClass = "hep-ph",
    doi = "10.1140/epjc/s10052-017-4996-1",
    journal = "Eur. Phys. J. C",
    volume = "77",
    number = "7",
    pages = "484",
    year = "2017"
}

@article{Burrows:1995ww,
    author = "Burrows, Adam and Hayes, John and Fryxell, Bruce A.",
    title = "{On the nature of core collapse supernova explosions}",
    eprint = "astro-ph/9506061",
    archivePrefix = "arXiv",
    reportNumber = "STEWARD-1236",
    doi = "10.1086/176188",
    journal = "Astrophys. J.",
    volume = "450",
    pages = "830",
    year = "1995"
}

@article{Kotake:2005zn,
    author = "Kotake, Kei and Sato, Katsuhiko and Takahashi, Keitaro",
    title = "{Explosion mechanism, neutrino burst, and gravitational wave in core-collapse supernovae}",
    eprint = "astro-ph/0509456",
    archivePrefix = "arXiv",
    doi = "10.1088/0034-4885/69/4/R03",
    journal = "Rept. Prog. Phys.",
    volume = "69",
    pages = "971--1144",
    year = "2006"
}

@article{Janka:2012wk,
    author = "Janka, Hans-Thomas",
    title = "{Explosion Mechanisms of Core-Collapse Supernovae}",
    eprint = "1206.2503",
    archivePrefix = "arXiv",
    primaryClass = "astro-ph.SR",
    doi = "10.1146/annurev-nucl-102711-094901",
    journal = "Ann. Rev. Nucl. Part. Sci.",
    volume = "62",
    pages = "407--451",
    year = "2012"
}

@article{Gossan:2015xda,
    author = "Gossan, S. E. and Sutton, P. and Stuver, A. and Zanolin, M. and Gill, K. and Ott, C. D.",
    title = "{Observing Gravitational Waves from Core-Collapse Supernovae in the Advanced Detector Era}",
    eprint = "1511.02836",
    archivePrefix = "arXiv",
    primaryClass = "astro-ph.HE",
    doi = "10.1103/PhysRevD.93.042002",
    journal = "Phys. Rev. D",
    volume = "93",
    number = "4",
    pages = "042002",
    year = "2016"
}

@article{Szczepanczyk:2021bka,
    author = "Szczepanczyk, Marek and others",
    title = "{Detecting and reconstructing gravitational waves from the next galactic core-collapse supernova in the advanced detector era}",
    eprint = "2104.06462",
    archivePrefix = "arXiv",
    primaryClass = "astro-ph.HE",
    doi = "10.1103/PhysRevD.104.102002",
    journal = "Phys. Rev. D",
    volume = "104",
    number = "10",
    pages = "102002",
    year = "2021"
}

@article{KAGRA:2021tnv,
    author = "Abbott, R. and others",
    collaboration = "LIGO, Virgo, KAGRA",
    title = "{All-sky search for short gravitational-wave bursts in the third Advanced LIGO and Advanced Virgo run}",
    eprint = "2107.03701",
    archivePrefix = "arXiv",
    primaryClass = "gr-qc",
    reportNumber = "P2100045",
    doi = "10.1103/PhysRevD.104.122004",
    journal = "Phys. Rev. D",
    volume = "104",
    number = "12",
    pages = "122004",
    year = "2021"
}

@article{Bergh:1991ek,
    author = "Bergh, S. V. and Tammann, G. A.",
    title = "{Galactic and extragalactic supernova rates}",
    doi = "10.1146/annurev.aa.29.090191.002051",
    journal = "Ann. Rev. Astron. Astrophys.",
    volume = "29",
    pages = "363--407",
    year = "1991"
}

@article{Cappellaro:1993ns,
    author = "Cappellaro, E. and Turatto, M. and Benetti and Tsvetkov, D. Yu. and Bartunov, O. S. and Makarova, I. N.",
    title = "{The rate of supernovae. 2. The selection effects and the frequencies per unit blue luminosity}",
    eprint = "astro-ph/9302017",
    archivePrefix = "arXiv",
    journal = "Astron. Astrophys.",
    volume = "273",
    pages = "383",
    year = "1993"
}

@article{Tammann:1994ev,
    author = "Tammann, G. A. and Loeffler, W. and Schroder, A.",
    title = "{The Galactic supernova rate}",
    doi = "10.1086/192002",
    journal = "Astrophys. J. Suppl.",
    volume = "92",
    pages = "487--493",
    year = "1994"
}

@article{Diehl:2006cf,
    author = "Diehl, Roland and others",
    title = "{Radioactive Al-26 and massive stars in the galaxy}",
    eprint = "astro-ph/0601015",
    archivePrefix = "arXiv",
    doi = "10.1038/nature04364",
    journal = "Nature",
    volume = "439",
    pages = "45--47",
    year = "2006"
}

@article{Li:2010kc,
    author = "Li, Weidong and others",
    title = "{Nearby Supernova Rates from the Lick Observatory Supernova Search. II. The Observed Luminosity Functions and Fractions of Supernovae in a Complete Sample}",
    eprint = "1006.4612",
    archivePrefix = "arXiv",
    primaryClass = "astro-ph.SR",
    doi = "10.1111/j.1365-2966.2011.18160.x",
    journal = "Mon. Not. Roy. Astron. Soc.",
    volume = "412",
    pages = "1441",
    year = "2011"
}

@article{Adams:2013ana,
    author = "Adams, Scott M. and Kochanek, C. S. and Beacom, John F. and Vagins, Mark R. and Stanek, K. Z.",
    title = "{Observing the Next Galactic Supernova}",
    eprint = "1306.0559",
    archivePrefix = "arXiv",
    primaryClass = "astro-ph.HE",
    doi = "10.1088/0004-637X/778/2/164",
    journal = "Astrophys. J.",
    volume = "778",
    pages = "164",
    year = "2013"
}

@article{LIGOScientific:2024jxh,
    author = "Abac, A. G. and others",
    collaboration = "LIGO, Virgo, KAGRA",
    title = "{Search for gravitational waves emitted from SN 2023ixf}",
    eprint = "2410.16565",
    archivePrefix = "arXiv",
    primaryClass = "astro-ph.HE",
    reportNumber = "LIGO-P2400125",
    month = "10",
    year = "2024"
}

@article{LVK-O3-SGWB,
    author = "Abbott, R. and others",
    collaboration = "LIGO, Virgo, KAGRA",
    title = "{Upper limits on the isotropic gravitational-wave background from Advanced LIGO and Advanced Virgo\textquoteright{}s third observing run}",
    eprint = "2101.12130",
    archivePrefix = "arXiv",
    primaryClass = "gr-qc",
    reportNumber = "LIGO-DCC-P2000314",
    doi = "10.1103/PhysRevD.104.022004",
    journal = "Phys. Rev. D",
    volume = "104",
    number = "2",
    pages = "022004",
    year = "2021"
}

@ARTICLE{Martynov2019,
       author = {{Martynov}, Denis and {Miao}, Haixing and {Yang}, Huan and {Vivanco}, Francisco Hernandez and {Thrane}, Eric and {Smith}, Rory and {Lasky}, Paul and {East}, William E. and {Adhikari}, Rana and {Bauswein}, Andreas and {Brooks}, Aidan and {Chen}, Yanbei and {Corbitt}, Thomas and {Freise}, Andreas and {Grote}, Hartmut and {Levin}, Yuri and {Zhao}, Chunnong and {Vecchio}, Alberto},
        title = "{Exploring the sensitivity of gravitational wave detectors to neutron star physics}",
      journal = {\prd},
         year = 2019,
        month = may,
       volume = {99},
       number = {10},
          eid = {102004},
        pages = {102004},
          doi = {10.1103/PhysRevD.99.102004},
archivePrefix = {arXiv},
       eprint = {1901.03885},
 primaryClass = {astro-ph.IM},
       adsurl = {https://ui.adsabs.harvard.edu/abs/2019PhRvD..99j2004M}
}

@article{Baldry:2003xi,
    author = "Baldry, Ivan K. and Glazebrook, Karl",
    title = "{Constraints on a universal IMF from UV to near-IR galaxy luminosity densities}",
    eprint = "astro-ph/0304423",
    archivePrefix = "arXiv",
    doi = "10.1086/376502",
    journal = "Astrophys. J.",
    volume = "593",
    pages = "258--271",
    year = "2003"
}

@article{Speagle_2020,
   title={dynesty: a dynamic nested sampling package for estimating Bayesian posteriors and evidences},
   volume={493},
   ISSN={1365-2966},
   url={http://dx.doi.org/10.1093/mnras/staa278},
   DOI={10.1093/mnras/staa278},
   number={3},
   journal={Monthly Notices of the Royal Astronomical Society},
   publisher={Oxford University Press (OUP)},
   author={Speagle, Joshua S},
   year={2020},
   month=feb, pages={3132–3158} }

@article{Ashton:2018jfp,
    author = "Ashton, Gregory and others",
    title = "{BILBY: A user-friendly Bayesian inference library for gravitational-wave astronomy}",
    eprint = "1811.02042",
    archivePrefix = "arXiv",
    primaryClass = "astro-ph.IM",
    doi = "10.3847/1538-4365/ab06fc",
    journal = "Astrophys. J. Suppl.",
    volume = "241",
    number = "2",
    pages = "27",
    year = "2019"
}

@TechReport{Ott2010,
  author = {Ott, C. D.},
  title  = {Tech. Rep. LIGO-T1000553-v2},
  year   = {2010},
  note   = {Accessed: 2023-10-01},
  school = {LIGO Scientific Collaboration},
  url    = {https://dcc.ligo.org/LIGO-T1000553-v2/public},
}

@Article{Piro2007,
  author        = {Piro, Anthony L. and Pfahl, Eric},
  journal       = {Astrophys. J.},
  title         = {{Fragmentation of Collapsar Disks and the Production of Gravitational Waves}},
  year          = {2007},
  pages         = {1173},
  volume        = {658},
  archiveprefix = {arXiv},
  doi           = {10.1086/511672},
  eprint        = {astro-ph/0610696},
}

@Article{Gupta2024,
  author        = {Gupta, Ish and others},
  journal       = {Class. Quant. Grav.},
  title         = {{Characterizing gravitational wave detector networks: from A$^\sharp$ to cosmic explorer}},
  year          = {2024},
  number        = {24},
  pages         = {245001},
  volume        = {41},
  archiveprefix = {arXiv},
  doi           = {10.1088/1361-6382/ad7b99},
  eprint        = {2307.10421},
  primaryclass  = {gr-qc},
  reportnumber  = {CE Document No. P2300019, CE Document No. P2300019-v2},
}

@Article{Andresen2019,
  author        = {Andresen, H. and M\"uller, E. and Janka, H. Th. and Summa, A. and Gill, K. and Zanolin, M.},
  journal       = {Mon. Not. Roy. Astron. Soc.},
  title         = {{Gravitational waves from 3D core-collapse supernova models: The impact of moderate progenitor rotation}},
  year          = {2019},
  number        = {2},
  pages         = {2238--2253},
  volume        = {486},
  archiveprefix = {arXiv},
  doi           = {10.1093/mnras/stz990},
  eprint        = {1810.07638},
  primaryclass  = {astro-ph.HE},
}

@Article{Abbott2020,
  author        = {Abbott, B. P. and others},
  journal       = {Phys. Rev. D},
  title         = {{Optically targeted search for gravitational waves emitted by core-collapse supernovae during the first and second observing runs of advanced LIGO and advanced Virgo}},
  year          = {2020},
  number        = {8},
  pages         = {084002},
  volume        = {101},
  archiveprefix = {arXiv},
  collaboration = {LIGO, Virgo},
  doi           = {10.1103/PhysRevD.101.084002},
  eprint        = {1908.03584},
  primaryclass  = {astro-ph.HE},
  reportnumber  = {LIGO-P1700177},
}

@article{Allen:2001ay,
    author = "Allen, Bruce and Creighton, Jolien D. E. and Flanagan, Eanna E. and Romano, Joseph D.",
    title = "{Robust statistics for deterministic and stochastic gravitational waves in nonGaussian noise. 1. Frequentist analyses}",
    eprint = "gr-qc/0105100",
    archivePrefix = "arXiv",
    doi = "10.1103/PhysRevD.65.122002",
    journal = "Phys. Rev. D",
    volume = "65",
    pages = "122002",
    year = "2002"
}

@article{Allen:1997ad,
    author = "Allen, Bruce and Romano, Joseph D.",
    title = "{Detecting a stochastic background of gravitational radiation: Signal processing strategies and sensitivities}",
    eprint = "gr-qc/9710117",
    archivePrefix = "arXiv",
    reportNumber = "WISC-MILW-97-TH-14",
    doi = "10.1103/PhysRevD.59.102001",
    journal = "Phys. Rev. D",
    volume = "59",
    pages = "102001",
    year = "1999"
}

@article{Flanagan:1997sx,
    author = "Flanagan, Eanna E. and Hughes, Scott A.",
    title = "{Measuring gravitational waves from binary black hole coalescences: 1. Signal-to-noise for inspiral, merger, and ringdown}",
    eprint = "gr-qc/9701039",
    archivePrefix = "arXiv",
    reportNumber = "GRP-456",
    doi = "10.1103/PhysRevD.57.4535",
    journal = "Phys. Rev. D",
    volume = "57",
    pages = "4535--4565",
    year = "1998"
}

@techreport{LIGO:2018instrument,
  author       = {{LIGO Scientific Collaboration}},
  title        = {Instrument Science White Paper 2018},
  institution  = {LIGO Document Control Center},
  number       = {T1800042-v4},
  year         = {2018},
  month        = mar,
  url          = {https://dcc.ligo.org/public/0149/T1800042/004/T1800042-v4.pdf},}

@article{Hild:2010id,
    author = "Hild, S. and others",
    title = "{Sensitivity Studies for Third-Generation Gravitational Wave Observatories}",
    eprint = "1012.0908",
    archivePrefix = "arXiv",
    primaryClass = "gr-qc",
    doi = "10.1088/0264-9381/28/9/094013",
    journal = "Class. Quant. Grav.",
    volume = "28",
    pages = "094013",
    year = "2011"
}

@article{Planck:2018vyg,
    author = "Aghanim, N. and others",
    collaboration = "Planck Collaboration",
    title = "{Planck 2018 results. VI. Cosmological parameters}",
    eprint = "1807.06209",
    archivePrefix = "arXiv",
    primaryClass = "astro-ph.CO",
    doi = "10.1051/0004-6361/201833910",
    journal = "Astron. Astrophys.",
    volume = "641",
    pages = "A6",
    year = "2020",
    note = "[Erratum: Astron.Astrophys. 652, C4 (2021)]"
}

@article{Madau:2014bja,
    author = "Madau, Piero and Dickinson, Mark",
    title = "{Cosmic Star Formation History}",
    eprint = "1403.0007",
    archivePrefix = "arXiv",
    primaryClass = "astro-ph.CO",
    doi = "10.1146/annurev-astro-081811-125615",
    journal = "Ann. Rev. Astron. Astrophys.",
    volume = "52",
    pages = "415--486",
    year = "2014"
}

@article{LIGOScientific:2025slb,
    author = "Abac, A. G. and others",
    collaboration = "LIGO, Virgo, KAGRA",
    title = "{GWTC-4.0: Updating the Gravitational-Wave Transient Catalog with Observations from the First Part of the Fourth LIGO-Virgo-KAGRA Observing Run}",
    eprint = "2508.18082",
    archivePrefix = "arXiv",
    primaryClass = "gr-qc",
    reportNumber = "LIGO-P2400386",
    month = "8",
    year = "2025"
}

@ARTICLE{LVK-O4a-sgwb,
       author = { {Abac}, A.~G. and others},
       collaboration = "LIGO, Virgo, KAGRA",
        title = "{Upper Limits on the Isotropic Gravitational-Wave Background from the first part of LIGO, Virgo, and KAGRA's fourth Observing Run}",
      journal = {arXiv e-prints},
         year = 2025,
        month = aug,
          eid = {arXiv:2508.20721},
        pages = {arXiv:2508.20721},
          doi = {10.48550/arXiv.2508.20721},
archivePrefix = {arXiv},
       eprint = {2508.20721},
 primaryClass = {gr-qc},
       adsurl = {https://ui.adsabs.harvard.edu/abs/2025arXiv250820721T}
}

@article{Evans:2021gyd,
    author = "Evans, Matthew and others",
    title = "{A Horizon Study for Cosmic Explorer: Science, Observatories, and Community}",
    eprint = "2109.09882",
    archivePrefix = "arXiv",
    primaryClass = "astro-ph.IM",
    reportNumber = "CE-P2100003-v7, Cosmic Explorer technical report CE-P2100003-v6",
    month = "9",
    year = "2021"
}

@article{LIGOScientific:2016wof,
    author = "Abbott, B. P. and others",
    collaboration = "LIGO",
    title = "{Exploring the Sensitivity of Next Generation Gravitational Wave Detectors}",
    eprint = "1607.08697",
    archivePrefix = "arXiv",
    primaryClass = "astro-ph.IM",
    reportNumber = "LIGO-P1600143",
    doi = "10.1088/1361-6382/aa51f4",
    journal = "Class. Quant. Grav.",
    volume = "34",
    number = "4",
    pages = "044001",
    year = "2017"
}

@article{LIGOScientific:2016jvu,
    author = "Abbott, B. P. and others",
    collaboration = "LIGO, Virgo",
    title = "{A First Targeted Search for Gravitational-Wave Bursts from Core-Collapse Supernovae in Data of First-Generation Laser Interferometer Detectors}",
    eprint = "1605.01785",
    archivePrefix = "arXiv",
    primaryClass = "gr-qc",
    reportNumber = "LIGO-P1400208",
    doi = "10.1103/PhysRevD.94.102001",
    journal = "Phys. Rev. D",
    volume = "94",
    number = "10",
    pages = "102001",
    year = "2016"
}

@article{LVK:O3opendata,
    author = "Abbott, R. and others",
    collaboration = "LIGO, Virgo, KAGRA",
    title = "{Open Data from the Third Observing Run of LIGO, Virgo, KAGRA, and GEO}",
    eprint = "2302.03676",
    archivePrefix = "arXiv",
    primaryClass = "gr-qc",
    reportNumber = "LIGO-P2200316",
    doi = "10.3847/1538-4365/acdc9f",
    journal = "Astrophys. J. Suppl.",
    volume = "267",
    number = "2",
    pages = "29",
    year = "2023"
}

@article{Vartanyan:2023sxm,
    author = "Vartanyan, David and Burrows, Adam and Wang, Tianshu and Coleman, Matthew S. B. and White, Christopher J.",
    title = "{Gravitational-wave signature of core-collapse supernovae}",
    eprint = "2302.07092",
    archivePrefix = "arXiv",
    primaryClass = "astro-ph.HE",
    doi = "10.1103/PhysRevD.107.103015",
    journal = "Phys. Rev. D",
    volume = "107",
    number = "10",
    pages = "103015",
    year = "2023"
}

@article{Choi:2024irp,
    author = "Choi, Lyla and Burrows, Adam and Vartanyan, David",
    title = "{Gravitational-wave and Gravitational-wave Memory Signatures of Core-collapse Supernovae}",
    eprint = "2408.01525",
    archivePrefix = "arXiv",
    primaryClass = "astro-ph.HE",
    doi = "10.3847/1538-4357/ad74f8",
    journal = "Astrophys. J.",
    volume = "975",
    number = "1",
    pages = "12",
    year = "2024",
    note = "[Erratum: Astrophys.J. 985, 268 (2025)]"
}

@article{Whelan:2012-calibration,
    author = "Whelan, John T. and Robinson, Emma L. and Romano, Joseph D. and Thrane, Eric H.",
    title = "{Treatment of Calibration Uncertainty in Multi-Baseline Cross-Correlation Searches for Gravitational Waves}",
    eprint = "1205.3112",
    archivePrefix = "arXiv",
    primaryClass = "gr-qc",
    reportNumber = "LIGO-P1200051, LIGO-P1200051-V4",
    doi = "10.1088/1742-6596/484/1/012027",
    journal = "J. Phys. Conf. Ser.",
    volume = "484",
    pages = "012027",
    year = "2014"
}

@article{Sun:2020wke-calibration,
    author = "Sun, Ling and others",
    title = "{Characterization of systematic error in Advanced LIGO calibration}",
    eprint = "2005.02531",
    archivePrefix = "arXiv",
    primaryClass = "astro-ph.IM",
    doi = "10.1088/1361-6382/abb14e",
    journal = "Class. Quant. Grav.",
    volume = "37",
    number = "22",
    pages = "225008",
    year = "2020"
}

@article{Romero:2021kby-calibration,
    author = "Romero, Alba and Martinovic, Katarina and Callister, Thomas A. and Guo, Huai-Ke and Mart{\'\i}nez, Mario and Sakellariadou, Mairi and Yang, Feng-Wei and Zhao, Yue",
    title = "{Implications for First-Order Cosmological Phase Transitions from the Third LIGO-Virgo Observing Run}",
    eprint = "2102.01714",
    archivePrefix = "arXiv",
    primaryClass = "hep-ph",
    doi = "10.1103/PhysRevLett.126.151301",
    journal = "Phys. Rev. Lett.",
    volume = "126",
    number = "15",
    pages = "151301",
    year = "2021"
}

@article{gwosc,
    eprint = "https://gwosc.org",
}

@article{data-available,
    author = "R. Abbott et al. (LIGO-Virgo-KAGRA Collaborations)",
    eprint = "https://dcc.ligo.org/G2001287/public",
}
\end{document}